\documentclass[12pt]{article}
\usepackage{amssymb,amsmath}
\usepackage{epsfig}
\newcommand{\sect}[1]{\setcounter{equation}{0}\section{#1}}
\renewcommand{\theequation}{\arabic{section}.\arabic{equation}}

\def\N{{\mathcal N}}
\def\L{{\mathcal L}}
\def\E{{\mathcal E}}
\def\S{{\mathcal S}}
\def\O{{\mathcal O}}
\def\D{{\mathcal D}}
\def\r{\rho}
\def\a{\alpha}
\def\b{\beta}
\def\d{\delta}
\def\g{\gamma}
\def\G{\Gamma}
\def\s{\sigma}
\def\t{\tau}
\def\x{\xi}
\def\z{\zeta}
\def\m{\mu}
\def\n{\nu}
\def\k{\kappa}
\def\f{\phi}
\def\o{\omega}
\def\e{\eta}
\def\ep{\varepsilon}
\def\l{\lambda}
\def\th{\theta}
\def\tz{\tilde\zeta}
\def\p{\partial}
\def\rb{\right}
\def\lb{\left}
\def\axs{AdS_5\times S^5}
\newcommand{\eq}[1]{\begin{equation} #1 \end{equation}}
\newcommand{\al}[1]{\begin{align} #1 \end{align}}
\newcommand{\ml}[1]{\begin{multline} #1 \end{multline}}

\begin{document}
\begin{center}
{\bf{\Large Semiclassical quantization of Rotating Strings in Pilch-Warner geometry} \\
\vspace*{.35cm}
}

\vspace*{1cm}
H. Dimov${}^{\ddag}$, V. Filev${}^{\dagger}$,
R.C. Rashkov${}^{\dagger}$\footnote{e-mail: rash@phys.uni-sofia.bg}

\ \\

and 

\ \\
K.S. Viswanathan${}^{\flat}$\footnote{e-mail: kviswana@sfu.ca}\\

\ \\
${}^{\dagger}$Department of Physics, Sofia University, 1164 Sofia,
Bulgaria 

\ \\

${}^{\ddag}$ Department of Mathematics, University of 
Chemical Technology and Metallurgy, 1756 Sofia, Bulgaria

\ \\

${}^{\flat}$ Department of Physics, Simon Fraser University \\
Burnaby, BC, V5A 1S6, Canada
\end{center}

\vspace*{.8cm}

\begin{abstract}
Some of the recent important developments in understanding string/ gauge dualities are based on the idea of highly symmetric motion of ``string solitons'' in $AdS_5\times S^5$ geometry originally suggested by Gubser, Klebanov and Polyakov. In this paper we study symmetric motion of certain string configurations in so called Pilch-Warner geometry. The two-form field $A_2$ breaks down the supersymmetry to $\mathcal{N}=1$ but for the string configurations considered in this paper the classical values of the energy and the spin are the same as for string in $AdS\times S^5$. Although trivial at classical level, the presence of NS-NS antisymmetric field couples the fluctuation modes that indicates changes in the quantum corrections to the energy spectrum. We compare our results with those obtained in the case of pp-wave limit in \cite{myers, pw}.
\end{abstract}

\vspace*{.8cm}

\section{Introduction}

The recent research efforts and developments in string theory in the last
years were focused on the understanding of string/gauge duality and
especially AdS/CFT correspondence. AdS/CFT correspondence is based on
the conjecture that type {\bf II B} string theory in $AdS_5\times S^5$
background with large number of fluxes turned on $S^5$ is dual to
four dimensional $\N=4$ supersymmetric Yang-Mills theory (SYM).
 Actually this is the best studied case where the spectrum of a
state of string theory on $\axs$ corresponds to the spectrum of single
trace operators  in the gauge theory. The great interest in
AdS/CFT correspondence is inspired by the simple reason that it can be
used to make predictions about $\N=4$ SYM at strong coupling.

Until recently this conjecture has been mostly investigated in the so called zero mode approximation. In this case it was found that the supergravity modes on $\axs$ are in one to one correspondence with the chiral operators in the gauge theory. However, this restriction means that the radius of the curvature is large and all $\alpha' R$  corrections are negligible ($R$ is the radius of $AdS$). Although important, these considerations are not enough to verify the correspondence in its full extent and to extract useful information about the gauge theory at strong coupling. It is important therefore to go beyond the supergravity limit and to consider at least  $\alpha'
R$ correction.

The recent progress in string propagation in pp-wave background attracted much interest \cite{b1, b2, b3}. This background possesses at least two very attractive features: it is maximally supersymmetric and at the tree level Green-Schwarz superstring is exactly solvable \cite{mets1, metsT1}. Since this background can be obtained by taking the Penrose limit \cite{penr} of $\axs$ geometry, the next natural step then is to ask about AdS/CFT correspondence in these geometries. In a remarkable paper, Berenstein, Maldacena and Nastase (BMN) \cite{bmn} have been able to connect the tree level string theory and the dual SYM theory in a beautiful way. Starting from the gauge theory side, BMN have  been able to identify particular string states with gauge invariant operators with large R-charge $J$, relating the energy of the string states to the dimension of the operators. It has been shown that these identifications are consistent with all planar contributions \cite{santam} and non-planar diagrams for large $J$ \cite{sem, others}.

Although this is a very important development, it describes however only one particular corner of the full range of gauge invariant operators in SYM corresponding to large $J$. It is desirable therefore to extend these results to wider class of operators. Some ideas about the extension of AdS/CFT correspondence beyond the supergravity approximation were actually suggested by Polyakov \cite{pol1} but until recently they wasn't quite explored.

In an important recent development, Gubser, Klebanov and Polyakov \cite{gkp} (GKP) suggested another way of going beyond the supergravity approximation. The main idea is to consider a particular configuration of closed string in $\axs$ background executing a highly symmetric motion. The theory of a generic string in this background is highly non-linear, but  semiclassical treatment will ensure the existence of globally conserved quantum numbers. AdS/CFT correspondence will allow us to relate these quantum numbers to the dimensions of particular gauge invariant operators in SYM. Gubser, Klebanov and Polyakov investigated three examples:

a) rotating ``string soliton'' on $S^5$ stretched along the radial direction $\r$ of AdS part. This case represents string states carrying large R-charge and GKP have been able to reproduce the results obtained from string theory in pp-wave backgrounds \cite{bmn}.

b) rotating ``string soliton'' on $AdS_5$ stretched along the radial direction $\r$. This motion represents string states carrying spin $S$. The corresponding gauge invariant operators suggested in \cite{gkp} are important for deep inelastic scattering amplitudes as discussed in \cite{kehag}.

c) strings spinning on $S^5$ and stretched along an angular direction at $\r=0$.

All these states represent highly excited string states and since the analysis is semiclassical, the spin $S$ is assumed quantized. Shortly after \cite{gkp}, interesting generalizations appeared. In \cite{fts, ts02, ts} a more general solution which includes the above cases was investigated. Beside the solution interpolating between the cases considered by GKP, the authors of \cite{fts, ts02, ts} have been able to find a general formula relating the energy $E$, spin $S$ and the R-charge $J$. In \cite{russo} a more general solution interpolating between all the above cases was suggested. Further progress has been made for the case of black hole $AdS$ geometry \cite{petkou1} and confining AdS/CFT backgrounds \cite{petkou2, alishah1}. The study of the operators with large R-charge and twist two have been initiated in \cite{zar, kru}. Further interesting developments of this approach can be found also in \cite{wadia,alishah2}. We would note also the solution for circular configurations suggested in \cite{minah}. The string soliton in this case can be interpreted as a pulsating string. Further developments exploring this approach to the problem can be found also in \cite{nunez, pons, oz}.

From the preceding discussion it is clear that the extension of the semiclassical analysis to wider class of moving strings in $\axs$ background is highly desirable. The purpose of this note is to consider a more general sigma model of closed strings moving in the above background. One of the possibilities is to include closed string with antisymmetric B-field turned on. There are several known backgrounds that are solutions to the string equations. One of them is so called Pilch-Warner (PW) geometry \cite{pw1}. We will consider the case of rotating strings in PW geometry and will analyze the relation between the conserved quantum numbers in the theory - in our case the energy $E$ and the spin $S$. We found that, for the configurations we investigated, the relations between the energy and the spin are the same as in the case of rotating strings in $\axs$ background. The presence of the non-trivial NS-NS two-form field manifests itself in considerations of the fluctuation modes defining the quantum corrections to the conserved quantities. 

The paper is organized as follows. In the next section we present a brief review of rotating ``string soliton'' in $\axs$ background. In Section 3, we give a short description of Pilch-Warner geometry and its basic properties, as well as we briefly comment on the dual conformal theory. Next, in analogy with \cite{gkp}, we find the relation between the energy $E$ and spin $S$. After that,we study the quadratic fluctuations around the classical solutions and give explicit calculations of the quantum corrections to the anomalous dimensions. We conclude with summary and comparison of our results with those of taking Penrose limit and discuss some further directions of development. To make the presentation more selfcontained we review the Penrose limit of Pilch-Warner geometry in Appendix A. 

\vspace*{.5cm}

\sect{Rotating strings in $\axs$}

In this section we will review the semiclassical analysis of rotating strings \cite{gkp} following mainly \cite{fts, ts02, ts}.

The idea is to consider rotating and boosted closed ``string soliton'' stretched in particular directions of $\axs$ background. The general form of the non-linear sigma model Lagrangian (bosonic part) can be written as:
\eq{\label{1.1}
\L_B=\frac 12\sqrt{-g}g^{\a\b}\left[G^{(AdS_5)}_{ab}\p_\a X^a\, \p_\b
X^b + G^{(S^5)}_{nm}\p_\a X^n\,\p_\b X^m\right] 
}
The supergravity solutions of $\axs$ in global coordinates has the form:
\ml{\label{1.2}
ds^2=-cosh^2\r\,dt^2+d\r^2+sinh^2\r\,d\Omega^2_{S^3} +d\psi^2_1 \\
+ cos^2\psi_1\left(d\psi^2_2 + cos^2\psi_2\,d\Omega^2_{\tilde S^3}\right)  
}
where,
\al{ \label{1.3} 
& d\Omega^2_{S^3}=d\b^2_1+ cos^2\b_1\left(d\b^2_2+ cos^2\b_2
\,d\b^2_3\right) \\
      \label{1.4}
& d\Omega^2_{\tilde S^3}=d\psi^2_3+ cos^2\psi_3\left(d\psi^2_4+ cos^2\psi_4
\,d\psi^2_5\right)       
}
(we have incorporated $R^2$ factor in the overall constant multiplying the action, i.e. $1/\a\to R^2/\a$). Looking for solutions with conserved energy and angular momentum, we assume that the motion is executed along $\hat\phi=\b_3$ direction on $AdS_5$ and $\theta=\psi_5$ direction on $S^5$. Solution for closed string folded onto itself can be found by making the following ansatz:
\al{ \label{1.5}
& t=\k\t;    & & \hat\phi=\o\t;  & & \theta=\n\t \\
& \r=\r(\s); & & \b_1=\b_2=0; & & \psi_i=0 \quad (i=1,\dots ,4) \notag
}
where $\o,\k$ and $\n$ are constants.

We note that one can always use the reparametrization invariance of the worldsheet to set the time coordinate of space-time $t$ proportional to the worldsheet proper time $\t$. This means that our rod-like string is rotating with constant angular velocity in the corresponding spherical parts of $AdS_5$ and $S^5$.

Using the ansatz (\ref{1.5}) and the Lagrangian (\ref{1.1}) one can easily find the classical string equations, which in this case actually reduces to the following equation for $\r(\s)$:
\eq{ \label{1.6}
\frac{d^2\r}{d\s^2}=\lb(\k^2-\o^2\rb)cosh\r\,\,sinh\r
}
and the Virasoro constraints take the form:
\eq{ \label{1.7}
{\r'}^2=\k^2 cosh^2\r -\o^2 sinh^2\r -\n^2.
}
One can obtain also the induced metric which turns out to be conformally flat as expected:
\eq{ \label{1.7i}
g_{\a\b}={\r'}^2\e_{\a\b},
}
where prime on $\r$ denotes derivative with respect to $\s$. It is straightforward to obtain the constants of motion of the theory under consideration\footnote{We follow the notations of \cite{gkp} and \cite{fts}}: 
\al{ \label{1.8}
& E=\sqrt{\l}\int\limits_0^{2\pi}\frac{d\s}{2\pi}\lb(-\frac{\p\L_B}
{\p\dot t}\rb)=\sqrt{\l}\k\int\limits_0^{2\pi}\frac{d\s}{2\pi}
cosh^2\r \\
     \label{1.9}
& S=\sqrt{\l}\int\limits_0^{2\pi}\frac{d\s}{2\pi}\lb(\frac{\p\L_B}
{\p\dot{\hat\phi}}\rb)=\sqrt{\l}\o\int\limits_0^{2\pi}\frac{d\s}{2\pi}
sinh^2\r \\
     \label{1.10}
&  J= \sqrt{\l}\int\limits_0^{2\pi}\frac{d\s}{2\pi}\lb(\frac{\p\L_B}
{\p\dot\theta}\rb) =\sqrt{\l}\n
}
where we denoted $R^2/\a'$ as $\sqrt{\l}$. An immediate but important relation following from (\ref{1.8}) and (\ref{1.9}) connect $E$ and $S$:
\eq{ \label{1.11}
E=\sqrt{\l}\k+\frac\k\o S
}
Assuming that $\r(\s)$ has a turning point at $\pi/2$ and using the periodicity condition\footnote{Since the closed string is folded onto itself $\r(\s)=\r(\s+\pi)$.}, one can relate the three free parameters $\k, \o$ and $\n$ as follows:
\eq{ \label{1.12}
coth^2\r_0=\frac{\o^2-\n^2}{\k^2-\n^2}=1+\e
}
where $\r_0$ is the maximal value of $\r(\s)$. A new parameter $\e$ introduced in (\ref{1.12}) is useful for studying the limiting cases of short strings ($\e$ large) and long strings ($\e$ small). Using the constraint (\ref{1.7}), the following expressions for the periodicity condition and the constants of motion can be found:
\al{ \label{1.13}
& \sqrt{\k^2-\n^2}=\frac 1{\sqrt{\e}}
\,\, F\lb(\frac 12,\frac 12,1;-\frac 1\e\rb), \\
     \label{1.14}   
& E=\frac{\sqrt{\l}\k}{\sqrt{\k^2-\n^2}}\,\frac 1{\sqrt{\e}}
\,\, F\lb(-\frac 12,\frac 12,1;-\frac 1\e\rb), \\
      \label{1.15}
& S=\frac{\sqrt{\l}\o}{\sqrt{\k^2-\n^2}}\,
\frac 1{2\e^{3/2}}
\,\, F\lb(\frac 12,\frac 12,2;-\frac 1\e\rb), \\
}
where $F(\a,\b;\g;z)$ is the hypergeometric function.

In the case of short strings $1/\e$ is very small. Straightforward approximations lead to the relation between the energy  $\E=E/\sqrt{\l}$ and the spin $\S=S/\sqrt{\l}$:
\eq{ \label{1.16}
\E\approx \sqrt{\n^2+\frac{2\S}{\sqrt{1+\n^2}}} +
\frac{\sqrt{\n^2+\frac{2\S}{\sqrt{1+\n^2}}}}
{\sqrt{1+\n^2+\frac{2\S}{\sqrt{1+\n^2}}}}\S
}
One can study several limiting cases. The simplest one is when $\n<<1$ and therefore $\S<<1$. In this case
\eq{ \notag
E^2\approx J^2+2\sqrt{\l}S
}
which can be interpreted as a string moving in flat space along a circle with angular momentum $J$ and rotating in a plane with spin $S$.

The second case is when the boost energy is much smaller than the rotational energy: $\n^2<<\S$. The expression (\ref{1.16}) then reduces to the flat space Regge trajectory \cite{gkp, fts}:
\eq{ \label{1.17}
\E\approx \sqrt{2\S} +\frac{\n}{2\sqrt{2\S}}
}
The last and most interesting case is when the boost energy is large, i.e. $\n>>1$. In this case one can find:
\eq{ \label{1.18}
E\approx J + S +\frac{\l S}{2 J^2} +\dots
} 
which coincides with the leading order terms of BMN formula:
\eq{ \label{1.19}
E=J+\sum\limits_{n=-\infty}^\infty\sqrt{1+\frac{\l n^2}{J^2}}N_n 
+\O\lb(\frac 1{\sqrt{\l}}\rb)
}

Similar analysis can be done in the case of long strings. In this case very interesting relations between the constants of motion were found. We give below the results of \cite{gkp, fts, ts}

a) when $\n<< ln\,1/\e$
$$
E\approx S+\frac{\sqrt{\l}}\pi\,\,ln\,\lb(S/\sqrt{\l}\rb) +\frac{\pi J^2}
{2\sqrt{\l}\,\,ln\,\lb(S/\sqrt{\l}\rb)}
$$

b) when $\n>> ln\,1/\e$
$$
E\approx S+ J+\frac\l{2\pi^2 J}\,\,ln^2\lb( S/J\rb)
$$

We conclude this section by noting that  more general solutions can be obtained by taking $\psi_1=\psi_1(\s)$ \cite{russo}. In this case the solutions interpolate smoothly between the cases described above.

We note that the same approach have been applied to the case of $M$-theory with $G_2$ holonomy \cite{nunez} where the logarithmic dependence of the energy on the spin $J$ for certain configurations was reproduced.

\vspace*{.5cm}

\sect{Rotating strings in Pilch-Warner geometry}

In this Section we apply the ideology developed in the previous Section to the case of Pilch-Warner geometry. First we briefly review the form and the properties of the geometry and its gauge dual theory. Next we consider rotating closed strings in this geometry and fluctuations around particular classical configurations. We compute also the quantum corrections to the energy and other conserved quantities.

\subsection{Basic facts about Pilch-Warner geometry}

In this subsection we will review the basic facts about Pilch-Warner geometry. We discuss briefly also on the dual conformal theory.

The so called Pilch-Warner geometry is obtained in \cite{pw1} as a solution of five-dimensional $\mathcal{N}=8$ gauged supergravity lifted to ten dimensions. This geometry interpolates between the standard maximally supersymmetric $\axs$ in the ultraviolet (UV) critical point and warped $\axs$ times squashed five-sphere in the IR. In this paper we will restrict ourself to the IR critical point. An interesting feature on the gravity side is that it preserves $1/8$ supersymmetry everywhere while at IR fixed point it is enhanced to $1/4$. On the SYM side the IR fixed point corresponds to large $N$ limit of the superconformal $\mathcal{N}=1$ theory of Leigh-Strassler \cite{lstr}.

To begin with, we discuss first the form of the metric describing so called Pilch-Warner geometry. It is warped $AdS_5$ space times squashed $S^5$ given by\footnote{We use the notations of \cite{myers}.}:
\eq{
ds^2= \frac{X^{1/2}\cosh\chi}{\rho}\lb(e^{2A}ds^2(\mathbb{M}^4)
+dr^2\rb)+ds_5^2, \label{2.1a}
}
\ml{
ds_5^2= L_0^2\frac{X^{1/2}sech\chi}{\rho^3}\lb[d\theta^2+
\frac{\rho^6\cos^2\theta}{X}(\s_1^2+\s_2^2) \right. \\
\qquad +\frac{\rho^{12}\sin^2(2\theta)}{4X^2} \lb(\s_3+\frac{(2-\rho^6)}{2\rho^6} d\f\rb)^2 \\
+\left.\frac{\rho^6\cosh^2\chi}{16X^2}(3-\cos(2\theta))^2
\lb(d\f+\frac{4\cos^2\theta}{3-\cos(2\theta)}\s_3\rb)^2\rb],
\label{2.2}
}
where we used
\eq{
X(r,\th)=\cos^2\th+\rho^6\sin^2\th.
\notag }
and the left-invariant $SU(2)$ one-forms $\s_i$ are given in (\ref{2.6}) below.
The global isometries of the metric can be readily read off from (\ref{2.2}) and it is $SU(2)\times U(1)_\b\times U(1)_{\phi}$ taught only a certain combinations of $\b$ and $\phi$ is preserved by the two-form fields. The IR is described by $r\to -\infty$ which means the following IR limit for the fields:
\eq{
\chi\to 2/\sqrt{3};\quad \rho\to2^{1/6};\quad A(r)\to\frac{2^{5/3} r}{3 L_0} \notag}
All these together gives in global coordinates:
\ml{
ds^2(IR)= \frac{3\sqrt{3}L_0^2}{8}\sqrt{3-\cos2\th}
\lb(-cosh^2\r \,d\t^2+d\r^2+sinh^2\r \,d\Omega_3^2\rb) \\
+ds_5^2(IR), \label{2.1} 
}
\ml{
ds_5^2(IR)= \frac{\sqrt{3}L_0^2}{4}\sqrt{3-\cos2\th}\lb[d\theta^2+
\frac{4\cos^2\theta}{3-\cos2\th}(\s_1^2+\s_2^2) \right. \\
\left.+\frac{4\sin^2(2\theta)}{(3-\cos2\th)^2}\s_3^2 
+\frac{2}{3}\lb(d\f+\frac{4\cos^2\theta}{3-\cos(2\theta)}\s_3\rb)^2\rb],
\label{2.3}
}
This geometry is supported by the following five- and two-form fields:
\al{
F_5(IR)=& -\frac{8}{9}\frac{2^{1/3}}{L_0}\,(1+\star)\,\epsilon(\mathbb{E}^4) \wedge dr, \label{2.4} \\
A_2(IR)=& Ae^{-2i\f}\frac{L_0^2\cos\th}{2}\lb(d\th -\frac{2i\sin2\th}{3-\cos2\th}\s_3\rb)\wedge(\s_1+i\s_2),
\label{2.5}
}
where the left invariant $SU(2)$ one-forms are given by
\al{
& \s_1=\frac{1}{2}\lb(\sin\b d\a-\sin\a \cos\b d\g\rb) \notag \\
& \s_2=\frac{1}{2}\lb(\cos\b d\a+\sin\a \sin\b d\g\rb) \label{2.6} \\
& \s_3=-\frac{1}{2}\lb(d\b+\cos\a d\g\rb) \notag 
}
normalized by $d\s_i=\frac{1}{2}\varepsilon_{ijk}d\s_j\wedge d\s_k$\footnote{Note that our normalization is that employed in \cite{myers}  with the corresponding numerical coefficients in the metric (\ref{2.3}) (cf. ref.\cite{myers}).}. The three-form field strength is $G_3=dA_2=F_3+iH_3$.

Pilch-Warner geometry actually represents "holographic renormalization group flow" from $\mathcal{N}=4$ Yang-Mills theory in UV to Leigh-Strassler theory in the $\mathcal{N}=1$ IR fixed point by a deformation of $\axs$ background. In the dual gauge theory, the $SU(2)$ doublet parametrized by
\al{
& \begin{pmatrix}
u^1\\
u^2
\end{pmatrix} =e^{-i\f/2}\cos\th\,g\,
\begin{pmatrix}1\\0\end{pmatrix}, & u^3=e^{-i\f}\sin\th \notag \\
& & \notag \\
& g=
\begin{pmatrix} v^1 & -\bar v^2\\ v^2 & \bar v^1\end{pmatrix} \in SU(2) \notag
}
corresponds to two massless chiral superfields $\Phi^1$ and
$\Phi^2$ while the singlet corresponds to a massive superfield $\Phi^3$. The relevant $\mathcal{N}=1$ deformation of $\mathcal{N}=4$ SYM is define by the modified superpotential 
\eq{
U=g\,tr\Phi^3[\Phi^1,\Phi^2]+\frac{1}{2}m\,tr(\Phi^3)^2.
\notag
}
The $\mathcal{R}$-symmetry in this case is a direct product of $SU(2)$ acting on $\Phi^1$ and $\Phi^2$, and $U(1)_R$. The $U(1)_R$ charges for the superfields $\Phi^{1,2,3}$ ensuring long distance conformal dynamics are correspondingly $(\frac{1}{2},\frac{1}{2},1)$ (for more details see \cite{lstr, johnson} and references therein). In the next subsection we will study the AdS/CFT correspondence beyond the supergravity limit.

\vspace*{.5cm}

\subsection{Rotating strings in Pilch-Warner geometry}

In this subsection we develop semiclassical analysis of a closed string in Pilch-Warner geometry (\ref{1.3}). First of all we will consider classical solutions of closed string and their conserved quantities such as energy, $\mathcal{R}$-charge and spin. To start with, let us consider an  ansatz for the embedding coordinates representing string configuration of folded string executing symmetric motion in the space-time defined in (\ref{2.3}). 

The classical configuration we will study is realized by the following ansatz:
\al{
& t=\k\t;               & & \b_3=\o\t;& & \b_{i\neq 3}=0; \notag \\
& \r=\r(\s)=\r(\s+2\pi);& & \th=0;    & & \a=0;\label{3.1} \\
& \f=\f(\t);            & & \b=\b(\t);& & \g=\g(\t). \notag
}
It is straightforward exercise to check that the above ansatz solves the string equations of motion providing
\eq{
\ddot{\b}+\ddot{\g}+\ddot{\f}=0 \label{3.2}
}
We assume that the "string soliton" defined in (\ref{3.1}) is rotating in $\b,\g$ and $\f$ directions with\footnote{We note that with the choice (\ref{3.1}) the metric gets degenerated and the three degrees of freedom $\b,\g$ and $\f$ are replaced by one: $\psi=\b+\g+\f$. The same happens in the case of taking Penrose limit in \cite{myers}.}:
\al{
& \b=c_\b\t, \quad & &\g=c_\g\t, \quad & \f=c_\f\t \notag \\
&\n\overset{def}{=}\frac{2}{3}(c_\b+c_\g+c_\f), \quad & & \psi\overset{def}{=}\b+\g+\f \label{3.3}
}
With the choice (\ref{3.1}) and (\ref{3.3}) the bosonic part of the Lagrangian becomes:
\eq{
\mathcal{L}_B=\frac{3\sqrt{3}}{4\sqrt{2}}L_0^2\lb(\k^2cosh^2\r+{\r'}^2 -\o^2sinh^2\r-\n^2\rb)
\label{3.4}
}
From (\ref{3.4}) it is obvious that the equations of motion for $\r(\s)$ and the Virasoro constraints have formally the same form as in the case of rotating strings in $\axs$ considered in the previous Section:
\al{ 
&\frac{d^2\r}{d\s^2}=\lb(\k^2-\o^2\rb)cosh\r\,\,sinh\r \notag \\
&{\r'}^2=\k^2 cosh^2\r -\o^2 sinh^2\r -\n^2. \label{3.6}
}
Therefore, all the conclusions concerning the classical values of the conserved quantities including the limiting cases of short and long strings remain unchanged (up to numerical coefficients). This is because for the chosen configuration the CS term vanishes and the Lagrangian takes formally the same form. The difference from the case studied in Sec 1. is that in our case we can have different momenta along the angles $\f,\b,\g$ which will correspond to different operators in SYM. We should note that the famous logarithmic dependence for long strings also take place but it is not {\it a priori} obvious that there will be no higher power logarithmic terms coming from the quantum corrections\footnote{The absence of higher powers of logarithmic corrections in the case of rotating strings in $\axs$ is discussed in \cite{ts}.}. We will return to this question in the Concluding section.

It is worth to note that for point-like string the Lagrangian becomes
\eq{
\mathcal{L}_B=\frac{3\sqrt{3}}{4\sqrt{2}}L_0^2\lb(-\dot\t^2+\dot\psi^2\rb)
\label{3.5}
}
and the result should be compared with those of taking Penrose limit \cite{myers} (see the discussion in the Concluding section and the Appendix).

\subsection{Quadratic fluctuations}

In this Section we study the quadratic fluctuations around the classical solutions of previous subsection. The aim is
to find the leading quantum corrections to the energy spectrum which determines the anomalous dimensions in SYM.

A standard method for studying small fluctuations around $\bar X^\m$ is the use of Riemann normal coordinates developed in \cite{gsw}. We
start with the sigma model string action:
\eq{ \label{f3.1}
S=-\frac 1{4\pi\a'}\int d^2\s\lb\{
\sqrt{g}g^{\a\b}G_{\m\n}\p_\a X^\m \p_\b X^\n +
\ep^{\a\b} B_{\m\n}\p_\a X^\m \p_b X^\n \rb\}.
}
The small fluctuations around $\bar X^\m$ generates the following transformations for the entries of (\ref{f3.1}) (see for instance \cite{gsw})  
\al{
& \p_\a X^\m=\p_\a\bar X^\m + D_\a\x^\m+\frac 13 R^\m_{\l\k\n}\x^\l
\x^\k\p_\a\bar X^\n+\cdots \notag \\
& G_{\m\n}(X)= G_{\m\n}(\bar X)-\frac 13 R_{\m\r\n\k}\x^\r\x^\k 
+ \cdots \notag \\
& B_{\m\n}(X)= B_{\m\n}(\bar X)+D_\r B_{\m\n}(\bar X)\x^\r+
\frac 12 D_\l D_\r B_{\m\n}(\bar X)\x^\l\x^\r - \notag \\
&\qquad \qquad - \frac 16 R^\l_{\r\m\k}(\bar X)B_{\l\n}(\bar X)\x^\r\x^\k
+\frac 16 R^\l_{\r\n\k}B_{\l\m}(\bar X)\x^\r\x^\k+\cdots,
\notag
}
where
\eq{ 
D_\a\x^\m=\p_\a\x^\m+\G^\m_{\l\n}\x^\l\p_\a\bar X^\n,  \notag
}
and $R^\m_{\n\r\l}$ is the usual Riemann tensor for the ambient space.
The quadratic part can be easily extracted from (\ref{f3.1}) 
\eq{ \label{f3.2}
S^{(2)}= S^{(2)}_G + S^{(2)}_B ,\notag
}
where
\al{
& S^{(2)}_G=-\frac 1{2\pi\a'}\int d^2\s\,\sqrt{g}g^{\a\b}
\lb[ G_{\m\n}(\bar X) D_\a\x^\m D_\b\x^\n + A_{\m\n;\a\b}\x^\m\x^\n
\rb] \label{f3.3} \\
& S^{(2)}_B = -\frac 1{2\pi\a'}\int d^2\s\, \ep^{\a\b} \lb[
\p_\a\bar X^\d H_{\d\m\n}\x^\n D_\b\x^\m +
\frac 12 D_\l H_{\d\m\n}\x^\l\x^\d\p_\a\bar X^\m\p_\b\bar X^\n
\rb] .\label{f3.4}
}
In (\ref{f3.3}) $ A_{\m\n;\a\b}$ is defined as\footnote{The
expressions for the AdS part of the metric are analogous to those in \cite{dgts} for the open superstring. In our case there are additional contributions from the squashed sphere.}:
\eq{ \label{f3.5} 
A_{\m\n;\a\b}=R_{\m\d\n\k}(\bar X)\p_\a\bar X^\d \p_\b\bar X^\k.
}
It is convenient to rewrite $S^{(2)}_G$ in Lorentz frame 
\eq{
S^{(2)}_G=-\frac 1{4\pi\a'}\int d^2\s \sqrt{g}
\lb[g^{\a\b}\e_{AB}\D_\a\x^A\D_\b\x^B+
\lb(g^{\a\b}R_{ACBD}Y^C_\a Y^D_\b\rb)\x^A\x^B\rb],
\label{f3.7}
}
where
\al{ \label{f3.7a}
&\D_\a\x^A=\p_a\x^A+\o^{AB}_\a\x^B\\
& Y^A_\a=E^A_\m\p_\a\bar X^\m .\notag
}
In (\ref{f3.7a}) the coefficients of the projected spin connection are 
\al{
& \o^A_{B\m}=E^A_\n\lb(\p_\m E^\n_B+\G^\n_{\m\l} E^\l_B\rb) \notag \\
& \o^{AB}_\a=\o^{AB}_\m\p_\a\bar X^\m . \notag 
}

Now we have to apply the above general scheme to the case of Pilch-Warner geometry (\ref{2.1}-\ref{2.5}) and string configuration (\ref{3.1}), (\ref{3.3}).
It is tedious but straightforward to calculate the components of the projected spin connection for our concrete configuration and the result is given as follows\footnote{We will use indices with tilde for $S_5$ part and without tilde for AdS part. The same will be applied to Lorentz indices denoted by numbers.}

a) for $AdS$ part
\eq{
\o_{\bar{0}}^{01}=\kappa~ sinh\r; \quad \o_{\bar{0}}^{14}=\o~ cosh\r.
\label{fl1}
}

b) for $S_5$ part
\eq{
\o_{\bar{0}}^{\tilde{0}\tilde{3}}=\frac{1}{3}\left(4c_\f+c_\b+c_\g\right); \quad
\o_{\bar{0}}^{\tilde{4}\tilde{2}}=\frac{1}{3}\left(2c_\f+2c_\b-c_\g\right).
\label{fl2}
}
The mass terms can be computed by using the following expressions

a) for $AdS$ part
\eq{
\frac{3\sqrt{6}}{8}\,L^2_0 R_{ACBD}=-\e_{AB}\e_{CD}+\e_{AD}\e_{CB}.
\label{fl2'}
}

b) for $S_5$ part, after projection on $\theta=\a=0$, the only non-vanishing components are
\al{
& R_{\tilde\m\f\tilde\n\f}\xi^{\tilde\m}\xi^{\tilde\n}=R_{\tilde\m\b\tilde\n\b}
\xi^{\tilde\m}\xi^{\tilde\n}=R_{\tilde\m\g\tilde\n\g}\xi^{\tilde\m}
\xi^{\tilde\n}=\notag \\
& R_{\tilde\m\f\tilde\n\b}\xi^{\tilde\m}\xi^{\tilde\n}=R_{\tilde\m\f\tilde\n\g}
\xi^{\tilde\m}\xi^{\tilde\n}=R_{\tilde\m\b\tilde\n\g}
\xi^{\tilde\m}\xi^{\tilde\n},
\label{fl3}
}
with
\eq{
R_{\tilde\m\f\tilde\n\f}e^{\tilde\m}_Ae^{\tilde\n}_B=\frac{13}{9}\left[\d^{\tilde 0}_A
\d^{\tilde 0}_B+\d^{\tilde 3}_A\d^{\tilde 3}_B\right]+\frac{4}{9}
\left[\d^{\tilde 2}_A\d^{\tilde 2}_B+\d^{\tilde 4}_A\d^{\tilde 4}_B\right].
\label{fl4}
}
We will use the above expressions to write the Lagrangian describing fluctuations in $S_5$ part.

c) for the non-vanishing mixed terms $R_{\m\tilde\m\n\tilde\n}$, $\m,\n\in AdS$, $\tilde\m, \tilde\n\in S_5$ we find only two non-vanishing terms
\al{
& R_{t\tilde\m t\tilde\n}\xi^{\tilde\m}\xi^{\tilde\n}=\frac{3\sqrt{6}L^2_0}{16}
cosh^2\r\left[\zeta_{\tilde 0}^2+\zeta_{\tilde 3}^2\right] \notag \\
& R_{\b_3\tilde\m \b_3\tilde\n}\xi^{\tilde\m}\xi^{\tilde\n}=-\frac{3\sqrt{6}L^2_0}{16}
sinh^2\r\left[\zeta_{\tilde 0}^2+\zeta_{\tilde 3}^2\right]. 
\label{fl6}
}

The final form of $\mathcal{L}_B$ describing quadratic fluctuations in Lorentz frame is
\eq{
\mathcal{L}_B=-\frac{4}{\sqrt{3}}(c_\f+c_\b+c_\g)~
(\p_1\zeta^{\tilde 0},\p_1\zeta^{\tilde 3})~\hat A(\d-\frac{\pi}{2}) 
\begin{pmatrix}
\zeta^{\tilde 4} \\
-\zeta^{\tilde 2}
\end{pmatrix},
\label{fl7}
}
where $\d=2\f+\b+\chi$ and $\hat A\in SU(2)$ is
\eq{
\hat A(\d-\frac{\pi}{2})=
\begin{pmatrix}
\cos(\d-\frac{\pi}{2}) & -\sin(\d-\frac{\pi}{2}) \\
\sin(\d-\frac{\pi}{2}) & \cos(\d-\frac{\pi}{2}) 
\end{pmatrix}.
\label{fl8}
}

We find for the Lagrangian describing fluctuating modes in $AdS$ part the expression
\ml{
\mathcal{L}_{AdS}=-\p_\a\tilde t\p^\a\tilde t+ \p_\a\tilde\r\p^\a\tilde\r+\p_\a\tilde\f\p^\a\tilde\f+
\p_\a\tilde\b_i\p^\a\tilde\b_i-m_t^2\tilde t^2+m_\r^2\tilde\r^2+m_\b^2\tilde\b_i^2\\
+4\tilde\r(\kappa~sinh\bar\r\p_0\tilde t-\o~cosh\bar\r\p_0\tilde \f),
\label{fl9}
}
with
\al{
& \tilde t=\zeta^0, & & \tilde\r=\zeta^1, & & \tilde\b_2=\zeta^3 \notag \\
& \tilde\f=\zeta^4, & & \tilde\b_1=\zeta^2 && \label{fl10} \\
& m_t^2=2\bar{\r'}^2+\n^2-\kappa^2, & & m_\f^2=2\bar{\r'}^2+\n^2-\o^2 \notag \\
& m_\r^2=2\bar{\r'}^2+2\n^2-\kappa^2-\o^2, & & m_\b^2=2\bar{\r'}^2+\n^2.
\label{fl11}
}

Let us turn now to the quadratic fluctuations in $S_5$ part. By making use of (\ref{fl4}) we find for $\mathcal{L}_{S_5}$ the following expression:
\al{
\mathcal{L}_{S_5}= & -\left[\p_0\z_{\tilde 5}+\frac{1}{3}(4c_\f+c_\b+c_\g)\z_{\tilde 3}
\right]^2-\frac{1}{3}\left[\p_0\z_{\tilde 3}-\frac{1}{3}(4c_\f+c_\b+c_\g)
\z_{\tilde 5}\right]^2 \notag \\
& -\left[\p_0\z_{\tilde 2}-\frac{1}{3}(2c_\f+2c_\b-c_\g)\z_{\tilde 4}\right]^2
-\left[\p_0\z_{\tilde 4}+\frac{1}{3}(2c_\f+2c_\b-c_\g)\z_{\tilde 2}\right]^2
\notag \\
& -(\p_0\z_{\tilde 1})^2+(\p_1\z_a)^2+\n^2\left[\z_{\tilde 2}^2+\z_{\tilde 4}^2\right]
+(\frac{3}{2}\bar{\r'}+4\n^2)\left[\z_{\tilde 5}^2+\z_{\tilde 3}^2\right].
\label{fl12}
}
To diagonalize the terms in the square brackets we perform the following rotations
\eq{
\begin{pmatrix}
\z_{\tilde 5}\\
\z_{\tilde 3}
\end{pmatrix}
=\hat A(\frac{4c_\f+c_\b+c_\g}{3}\t)
\begin{pmatrix}
\tilde\z_{\tilde 5}\\
\tilde\z_{\tilde 3}
\end{pmatrix},
\label{fl13a}
}
\eq{
\begin{pmatrix}
\z_{\tilde 4}\\
\z_{\tilde 2}
\end{pmatrix}
=\hat A(\frac{2c_\f+2c_\b-c_\g}{3}\t)
\begin{pmatrix}
\tilde\z_{\tilde 4}\\
\tilde\z_{\tilde 2}
\end{pmatrix},
\label{fl13b}
}
which leaves the rest of $\mathcal{L}_{S_5}$ unchanged. The final from is given by
\eq{
\mathcal{L}_{S_5}=\p_\a\tilde\z_a\p^\a\tilde\z_a+(4M^2+\frac{3}{2}\bar{\r'}^2)
\left[\tilde\z_{\tilde 5}^2+\tilde\z_{\tilde 3}^2\right]+
M^2\left[\tilde\z_{\tilde 2}^2+\tilde\z_{\tilde 4}^2\right],
\label{fl14}
}
where we defined
\eq{
M^2=\n^2=\frac{4}{9}(c_\f+c_\b+c_\g)^2.
\label{15} 
}

The transformations (\ref{fl13a}),(\ref{fl13b}) however affect $\mathcal{L}_B$ since B-field is coupled to $S_5$ part. Choosing the arbitrary constant $\chi=-\pi/2$ as in \cite{pw}, the expression for $\mathcal{L}_B$ actually simplifies
\eq{
\mathcal{L}_B=2\sqrt{3}M\left[\p_1\tilde\z_{\tilde 5}\tilde\z_{\tilde 4}
-\p_1\tilde\z_{\tilde 3}\tilde\z_{\tilde 2}\right].
\label{fl16}
}
Adding (\ref{fl16}) to $\mathcal{L}_{S_5}$, we find for quadratic fluctuations in $S_5$ part the expression
\ml{
\mathcal{L}_{S_5}^{tot}=\p_\a\tilde\z^A\p^\a\tilde\z^A+
M^2\left[\tilde\z_{\tilde 2}^2+\tilde\z_{\tilde 4}^2\right]+
(4M^2+\frac{3}{2}\bar{\r'}^2)\left[\tilde\z_{\tilde 3}^2+\tilde\z_{\tilde 5}^2\right] \\
+2\sqrt{3}M\left(\tilde\z_{\tilde 4}\p_1\tilde\z_{\tilde 5}-
\tilde\z_{\tilde 2}\p_1\tilde\z_{\tilde 3}\right).
\label{fl17}
}

It is straightforward now to write the equations of motion:

a) for $S_5$ part
\al{
&\nabla^2 \tz_{\tilde 1}=0 \notag \\
&\nabla^2 \tz_{\tilde 2}-M^2\tz_{\tilde 4}+\sqrt{3}M~\p_1\tz_{\tilde 3}=0 \notag \\
&\nabla^2 \tz_{\tilde 3}-(4M^2+\frac{3}{2}\bar{\r'}^2)\tz_{\tilde 3}
-\sqrt{3}M~\p_1\tz_{\tilde 2}=0 \notag \\
&\nabla^2 \tz_{\tilde 4}-M^2\tz_{\tilde 4}-\sqrt{3}M~\p_1 \tz_{\tilde 5}=0 \label{fl20} \\
&\nabla^2 \tz_{\tilde 5}-(4M^2+\frac{3}{2}\bar{\r'}^2)\tz_{\tilde 5}+\sqrt{3}M~\p_1 \tz_{\tilde 4}=0 \notag 
}

b) for $AdS$ part
\al{
& \nabla^2\tilde t-(2\bar{\r'}^2+\n^2-\kappa^2)\tilde t+2\kappa~sinh\bar\r~\p_0 \tilde\r=0 \notag \\
& \nabla^2 \tilde\r-(2\bar{\r'}^2+2\n^2-\kappa^2-\o^2)\tilde\r
-2(\kappa~sinh\bar\r~\p_0\tilde t-\o~cosh\bar\r~\p_0\tilde\f)=0 \notag \\
& \nabla^2 \tilde\f-(2\bar{\r'}^2+\n^2-\o^2)\tilde\f-2\o~cosh\bar\r~\p_0\tilde\r=0 
\notag \\
& \nabla^2 \tilde\b_i-(2\bar{\r'}^2+\n^2)\tilde\b_i=0 .
\label{fl21}
}

Now we are in a position to calculate the contribution of the massive modes to the classical energy. To do that we will use the approximations in the two limiting cases --  short and long strings. The problem is that in eqs. (\ref{fl20}) and (\ref{fl21}) the mass terms contain $\s$-dependent part through ${\bar\r'}$ which can be accounted for by using long and short string approximations.

Let us consider first the case of short strings. In this case an approximate solution for $\bar{\r'}^2(\s)$ is
\eq{
\bar{\r'}^2\approx \frac{\cos^2\s}{\e} \notag
}
It is clear that the contribution from $\bar{\r'}^2$ is very small ($\e$ is very large) and it can be replaced by its average $<\cos^2\s>=1/2$ and the mass terms containing $\bar{\r'}^2$ will be corrected by terms $\sim 1/2\e$. The analysis of the contributions from $AdS_5$ part follows as in \cite{fts}. The contribution of $\b$-fields can be calculated by solving the corresponding eigenvalue problem and the solution is given by
\al{
& \b_i=\sum_{n}\b_n^i\exp{(-i\o_n+in\s)} \notag \\
& \o_n^2\approx n^2+\n^2+\frac{1}{\e}\label{fl30}
}
Due to the choice $t=\k\t$ the space-time energy is proportional to the worldsheet one and we can write

a) for small $\n$, $\n^2\ll\mathcal{S}\ll 1$
\eq{
E\approx\sqrt{\l}\lb(\sqrt{2\mathcal{S}}+\frac{\n^2}{2\sqrt{2\mathcal{S}}}\rb)+
\frac{1}{\sqrt{2\mathcal{S}}}\sum\limits_{n=-\infty}^\infty\lb(1-
\frac{\n^2}{4\mathcal{S}n^2}+\frac{\mathcal{S}}{n^2}\rb)|n|N_n+\cdots
\label{fl31}
}

b) for $\n\gg 1$
\eq{
E\approx \l\lb(\n+\mathcal{S}+\frac{\mathcal{S}}{2\n^2}\rb)+
\sum\limits_{n=-\infty}^\infty\lb(1+\frac{n^2}{2\n^2}-\frac{n^2\mathcal{S}}{\n^5}\rb)N_n
+\cdots
\label{fl32}
}
Taking into account that $2\bar{\r'}^2\approx 1/\e$, $\k^2-\n^2\approx 1/\e$ and $\o^2-\k^2\approx 1$ the equations for $t$-fluctuations becomes trivial, $\nabla^2 t=0$ while the frequencies for $\r$ and $\varphi$ are
\ml{
(\o_n^\pm)^2=n^2-1+\frac{1}{2}\lb[4\o^2(1+\frac{1}{4\e})^2-\frac{1}{\e}\pm \right.\\
\left. \sqrt{[4\o^2(1+\frac{1}{4\e})^2-\frac{1}{\e}]^2+16\o^2(1+\frac{1}{4\e})^2(n^2-1)}~\rb],
\label{fl33}
}
where $\o^2\approx 1+1/\e+\n^2$. Let us turn now to the more involving $S^5$ part. The mode expansion $\z^{\tilde i}=\sum_n\tz_n^{\tilde i}\exp(-i\o_n+in\s)$ and the approximation $\bar{\r'}^2\approx 1/(2\e)$ gives the following equations
\al{
& \lb[-\o_n^2+n^2+M^2\rb]\tz_n^{\tilde 2}-in\sqrt{3}M\tz_n^{\tilde 3}=0 \notag \\
& \lb[-\o_n^2+n^2+4M^2+\frac{3}{4\e}\rb]\tz_n^{\tilde 3}+in\sqrt{3}M\tz_n^{\tilde 2}=0 \notag \\
& \lb[-\o_n^2+n^2+M^2\rb]\tz_n^{\tilde 4}+in\sqrt{3}M\tz_n^{\tilde 5}=0 \label{fl34} \\
& \lb[-\o_n^2+n^2+4M^2+\frac{3}{4\e}\rb]\tz_n^{\tilde 5}-in\sqrt{3}M\tz_n^{\tilde 4}=0, \notag
}
where $M^2=\n^2$. The solutions for the frequencies are
\al{
& (\o_n^\pm)^2=\frac{1}{2}\lb[5M^2+2n^2+\frac{3}{4\e}\pm 
\sqrt{12n^2M^2+9(M^2+\frac{1}{4\e})^2}\rb], \notag \\
& \o_0^+=\sqrt{4M^2+\frac{3}{4\e}}, \qquad \o_0^-=M.
\label{fl34a}
}
We find for the mode expansion of the fluctuations the expressions
\ml{
\z^A(\t,\s)=\cos\sqrt{4M^2+\frac{3}{4\e}}\t\z_0^A+\frac{\a'}{\sqrt{4M^2 +\frac{3}{4\e}}}
\sin\sqrt{4M^2+\frac{3}{4\e}}\t p_0^A  \\
+i\sqrt{\frac{\a'}{2}}\sum\limits_{n\neq 0}
\lb[\frac{1}{\o_n^+}\lb(\z_n^Ae^{in\s}+\tz_n^Ae^{-in\s}\rb)e^{-i\o_n^+\t} \right. \\
\left.+\frac{1}{\o_n^-}\lb(\z_n^{A+2}e^{in\s}+\tz_n^{A+2}e^{-in\s}\rb)e^{-i\o_n^-\t}\rb]
\label{35}
}
\ml{
\z^{A+2}(\t,\s)=\cos M\t\z_0^{A+2}+\frac{\a'}{M}
\sin M\t p_0^{A+2} \\
+i\sqrt{\frac{\a'}{2}}\sum\limits_{n\neq 0}
\lb[\frac{c_n^+}{\o_n^+}\lb(\z_n^Ae^{in\s}-\tz_n^Ae^{-in\s}\rb)e^{-i\o_n^+\t} \right. \\
\left. +\frac{c_n^-}{\o_n^-}\lb(\z_n^{A+2}e^{in\s}-\tz_n^{A+2}e^{-in\s}\rb)e^{-i\o_n^-\t}\rb]
\label{fl36}
}
where
\eq{
c_n^\pm=\frac{i}{2n\sqrt{3}M}\lb(-3M-\frac{3}{4\e}\pm \sqrt{12n^2M^2+(3M^2+\frac{3}{4\e})^2}\rb); \quad c_n^+c_n^-=1.
\label{37}
}
To write the Hamiltonian in canonical form we must first normalize the oscillators so that they satisfy the canonical oscillator algebra. This can be achieved by rescaling the oscillators as follows
\al{
& \z_n^A=\sqrt{\o_n^+\frac{c_n^-}{c_n^--c_n^+}}\hat\z_n^A; && 
\z_{-n}^A=\sqrt{\o_n^+\frac{c_n^-}{c_n^--c_n^+}}\bar{\hat\z}_{n}^A \notag \\
& \z_n^{A+2}=\sqrt{\o_n^-\frac{c_n^+}{c_n^+-c_n^-}}\hat\z_n^{A+2}; && 
\z_{-n}^{A+2}=\sqrt{\o_n^-\frac{c_n^+}{c_n^+-c_n^-}}\bar{\hat\z}_{n}^{A+2}
\label{fl38}
}
and similarly for $\tz_n^A$. We also redefine the zero modes as
\al{
& \hat\z_0^A=\frac{1}{\sqrt{2\a'}}\frac{1}{(4m^2+\frac{3}{4\e})^{\frac{1}{4}}}
\lb(\a'p_0^A-\sqrt{4M^2+\frac{3}{4\e}}\z_0^A\rb) \notag\\
& \bar{\hat\z}_0^A=\frac{1}{\sqrt{2\a'}}\frac{1}{(4m^2+\frac{3}{4\e})^{\frac{1}{4}}}
\lb(\a'p_0^A+\sqrt{4M^2+\frac{3}{4\e}}\z_0^A\rb) \notag \\
& \hat\z_0^{A+2}=\frac{1}{\sqrt{2\a'M}}
\lb(\a'p_0^A-M\z_0^{A+2}\rb) \notag \\
& \bar{\hat\z}_0^{A+2}=\frac{1}{\sqrt{2\a'M}}
\lb(\a'p_0^{A+2}+M\z_0^A\rb) 
\label{fl38a}
}
The commutation relations for this oscillator basis are
\eq{
\lb[\hat\z_0^A,\bar{\hat\z}_0^B\rb]=\d^{AB};\quad
\lb[\hat\z_n^A,\bar{\hat\z}_m^B\rb]=\d_{nm}\d^{AB},
\label{fl39}
}
and the Hamiltonian takes the standard form
\ml{
H=E+\sqrt{4M^2+\frac{3}{4\e}}\sum\limits_{A=1,2}N_0^{(A)}+M\sum\limits_{A=3,4}N_0^{(A)}\\ +\sum\limits_{n>0}\lb[\o_n^+N_n^{(A)}+\o_n^-N_n^{(A+2)}\rb].
\label{fl40}
}
The occupation number operators are defined as usually by
\eq{
N_n^{(A)}=\bar{\hat\z}_n^A\hat\z_n^A+\bar{\tilde\z}_n^A\tilde\z_n^A;
\quad N_0^{(A)}=\bar{\hat\z}_0^A\hat\z_0^A.
\label{fl41}
}
It is straightforward now to expand the frequencies by making use that $1/\e$ is small.
Since $1/\e\approx 2\mathcal{S}/\sqrt{\n^2+1}$, the expansion of (\ref{fl34a}) can be written as
\eq{
\o_n^\pm\approx\o_n^{(0)\pm}+\frac{3}{8}\lb[1\pm \frac{3\n^2}{\sqrt{12n^2\n^2+9\n^4}}\rb]
\frac{1}{\sqrt{1+\n^2}}\frac{\mathcal{S}}{\o_n^{(0)\pm}},
\label{short1}
}
where
\eq{
\o_n^{(0)\pm}=sign(n)\frac{1}{\sqrt{2}}\lb[5\n^2+2n^2\pm\sqrt{12n^2\n^2+9\n^4}\rb]^{1/2}.
\label{short2}
}
In the case of $\n>> 1$ we get
\al{
& \o_n^+=2\n+\frac{n^2}{2\n}+\frac{3}{8\n^2}\mathcal{S} &&
\o_n^-=\n+\frac{n^4}{3\n^3}+\frac{n^2}{4\n^4}\mathcal{S} \notag \\
& \o_0^+=2\n+\frac{3}{8\n^2}\mathcal{S} && \o_0^-=\n .
\label{short3}
}
When $\n<<1$ the result is
\al{
& \o_n^\pm=n\lb(1\pm\frac{\sqrt{3}}{2}\frac{\n}{|n|}\rb)+\frac{3}{8n}\mathcal{S} \notag \\
& \o_0^+=\lb(2-\frac{3}{16}\mathcal{S}\rb)\n+\frac{3}{8\n}\mathcal{S}, \qquad
\o_0^-=\n .
\label{short4}
}
We note that 2d energy is related to the space-time energy by multiplicative factor $\sqrt{\k}$ where $\k^2\approx 1/\e+\n^2$.

The case of long string can be treated analogously. The only difference comes from the approximation of the $\s$-dependent term entering the masses.

Since $\e$ is very small, the situations is more tricky and the only nontrivial Virasoro constraint is
\al{
{\r'}^2 &=(\k^2-\n^2)\lb[1-\e~ sinh^2\r\rb] \notag \\
&=\frac{(\k^2-\n^2)}{2}\lb[2+\e-\e~ cosh~2\r\rb].
\label{long1}
}
One can integrate the equation (\ref{long1}) explicitly and the result is
\eq{
2\sqrt{\k^2-\n^2}\s=\lb[F\lb(\arcsin\sqrt{\frac{2+\e-\e~ cosh\r}{2}},\frac{1}{\sqrt{1+\e}}\rb)-K\lb(\frac{1}{\sqrt{1+\e}}\rb) \rb],
\label{long2}
}
where $F(\varphi,k)$ is the incomplete elliptic integral of first kind and $K(k)$ is the complete elliptic integral of first kind, $F(\pi/2,k)$. Since $\e$ is small one can expand around $k\approx 1$. After long but straightforward calculations one can find the following leading in $1/\e$ term for $sinh^2\r$ ($0\leq\s<\pi/2$)
\eq{
sinh^2\r\sim \e^{-\frac{2\s}{\pi}}.
\notag
}
The substitution in ${\r'}^2$ gives
\eq{
{\r'}^2=(\k^2-\n^2)\lb(1-\e~sinh^2\r\rb)\approx
(\k^2-\n^2)\lb(1-\e^{\frac{\pi-2\s}{\pi}}\rb).
\label{long3}
}
Taking into account that for long string
{\eq{
\k^2-\n^2\approx\frac{1}{\pi^2}\log^2\frac{1}{\e},
\notag
}
we get
\al{
{\r'}^2 &\approx \frac{1}{\pi^2}\log^2\frac{1}{\e}-
\frac{1}{\pi^2}\e^{\frac{\pi-2\s}{\pi}}\log^2\frac{1}{\e} \notag \\
&\approx\frac{1}{\pi^2}\log^2\e,
\label{long4}
}
where in the second line we used that $\e^\epsilon\log^2\frac{1}{\e}$, ($\epsilon>0$) vanishes when $\e\to 0$ and $\log^21/\e=\log^2\e$.

We note that the masses in this case are extremely heavy, but one can still evaluate the quantum corrections to the energy $E$. Repeating the steps in the analysis of short strings we find for the Hamiltonian 
\ml{
H=E+\sqrt{4M^2+\frac{3}{2\pi^2}\log^2\e}\sum\limits_{A=1,2}N_0^{(A)}+M\sum\limits_{A=3,4}N_0^{(A)}\\ +\sum\limits_{n>0}\lb[\o_n^+N_n^{(A)}+\o_n^-N_n^{(A+2)}\rb],
\label{long5}
}
where the frequencies are given by
\ml{
(\o_n^\pm)^2=\frac{1}{2}\lb[5M^2+2n^2+\frac{3}{2\pi^2}\log^2\e\right. \\
\left. \pm \sqrt{12n^2M^2+9(M^2+\frac{1}{2\pi^2}\log^2\e)^2}\rb]
\label{long6}
}
\eq{
\o_0^+=\sqrt{4M^2+\frac{3}{2\pi^2}\log^2\e}, \qquad \o_0^-=M.
\label{long7}
}
For long strings we can also expand the low frequencies in two limiting cases. When $\n^2<<(1/\pi^2)\log^21/\e$ we have $1/\e\approx \pi\mathcal{S}/2$ and the expressions for the low frequencies becomes
\al{
& \o_n^+\approx \frac{1}{\pi}\sqrt{\frac{3}{2}}\,\log\mathcal{S}+2\pi\n^2 \sqrt{\frac{2}{3}}\,
\frac{1}{\log\mathcal{S}}+(3n^2-4\n^2)\frac{2^{1/2}\pi^3\n^2}{3^{3/2}\log^3\mathcal{S}} \notag \\
& \o_n^-\approx \n\lb(1-\frac{\pi^2n^2}{\log^2\mathcal{S}}\rb) \label{short5} \\
& \o_0^+\approx \sqrt{\frac{3}{2}}\,\frac{\log\mathcal{S}}{\pi}+2\n^2\pi
\sqrt{\frac{2}{3}}\,\frac{1}{\log\mathcal{S}}-
4\frac{2^{1/2}\pi^3\n^4}{3^{3/2}\log^3\mathcal{S}}, 
\quad \o_0^-=\n.
\notag 
}
Analogous expansion for low frequencies can be applied also in the other limiting case -- $\n^2>>(1/\pi^2)\log^21/\e$. To the leading order we find
\al{
& \o_n^+\approx 2\n+\frac{3\log^2(2\mathcal{S}/\n)+4n^2\pi^2}{8\pi^2\n}, \notag \\
& \o_n^-\approx \n+\frac{n^2[3\log^2(2\mathcal{S}/\n)+2n^2\pi^2]}{12\pi^2\n^3},
\label{long8} \\
& \o_0^+2\n+\frac{3}{8\pi^2\n}\log^2(2\mathcal{S}/\n), \quad \o_0^-=\n
\notag
}

We note that in the case of long strings the multiplicative factor relating world-sheet and space-time energy is given by $\k^2\approx \n^2+(1/\pi^2)\log^21/\e$.

\vspace*{.5cm}
 
\sect{Conclusions}

In this paper we considered the semiclassical rotating string in Pilch-Warner background. For the classical configuration we considered we found expression for the classical energy and the corresponding quantum corrections in the case of short and long strings. The classical analysis for our configuration follows closely that in $\axs$ case and we found the same qualitative picture. This is because for our configuration Chern-Simons term vanishes and the string do not feel the presence of the non-trivial B-field. In the case of short string we recovered to the leading order the familiar Regge behavior and gave explicit expression for the classical energy. In the case of long strings we obtained the famous logarithmic dependence of the energy in $\mathcal{S}$.

In Section 3 the quantum fluctuations around our classical string configuration is considered. We calculated the quantum corrections to the energy in the two limiting cases -- short and long strings. The presence of B-field now significantly affects the solutions because it couples the fluctuation modes in $S^5$ part in a nontrivial way. We calculated the frequencies that enters the 2d Hamiltonian for various cases of short and long strings.

Although it is well known that expanding near null geodesic one gets  quadratic action that is the same as the full string action in the corresponding plane wave geometry obtained by Penrose limit, certainly it is interesting to compare our results with those obtained by taking pp-wave limit \cite{myers,pw}\footnote{We thank A. Tseytlin for comments on this point.}. To do that, we consider the following special case:
\eq{
\o=\kappa=\n.
\label{fl22}
}
For this choice of the parameters all the fields in AdS part becomes massless (see (\ref{3.6}) and (\ref{fl11}))
\eq{
m_t^2=m_\r^2=m_\f^2=0.
\label{fl23}
}
To diagonalize the last term in (\ref{fl9}) we perform a boost defined by
\al{
&\tilde \f=cosh\bar\r~\f-sinh\bar\r~t \notag \\
&\tilde t=cosh\bar\r~t-sinh\bar\r~\f.
\label{fl24}
}
The boost (\ref{fl24}) combined with the $SO(2)$ rotation
\eq{
\begin{pmatrix}
\hat\r\\
\hat\f
\end{pmatrix}
=\hat A(\n\t)
\begin{pmatrix}
\r\\
\tilde\f
\end{pmatrix},
\label{fl25}
}
brings the Lagrangian to the form
\eq{
\mathcal{L}_{AdS}=-\p_\a\tilde t\p^\a\tilde t+\p_\a\hat\f\p^\a\hat\r+\p_\a\b_i\p^\a\b_i  +\n^2\left[\hat\r^2+\hat\f^2+\b_1^2+\b_2^2\right].
\label{fl26}
}
To keep the notations close to those in \cite{myers} we rename the fields as follows
\al{
&t=Z, && \hat\r=X_6, && \b_2=X_8 \notag \\
&\hat\f=X_5, && \b_1=X_7,
\label{fl27}
}
and the Lagrangian in these notations becomes
\eq{
\mathcal{L}_{AdS}=-\p_\a Z\p^\a Z+\p_\a X^p\p^\a X^p+M^2X^p,
\label{fl28}
}
where we used that $\n=M$ and $p=5,6,7,8$. 

The resulting equations of motion for AdS part are
\al{
& \nabla^2 Z=0 \notag \\
& \nabla^2 X^p -M^2X^p=0.
\label{fl29}
}
To compare with the case of pp-wave limit we simply choose point-like string configuration $\r=\r_0$ with $\r'=0$ (which, for our choice of parameters (\ref{fl22}), follows from (\ref{3.6}) and (\ref{fl11}) and therefore $\r$ is $\s$-independent). The complete set of equations (\ref{fl20}) and (\ref{fl29}) reduces to those given in the Appendix, see (\ref{a.9}).\footnote{See also \cite{myers}.} The analysis then proceeds as in the Appendix with the corresponding corrections to the anomalous dimensions of the operators in SYM. Therefore, we can conclude that point-like string limit of our semiclassical rotating string configuration in PW geometry reproduces the results obtained by taking pp-wave limit of the corresponding supersting theory.

To conclude, we note that the considerations in this paper were restricted to the bosonic sector of the theory. It would be interesting to proceed with the analysis of the fermionic part, the supersymmetry and more detailed analysis of AdS/CFT correspondence. We hope to return to this point in the near future.
\vspace*{.8cm}

{{\large{\bf Acknowledgements:}} R.R. would like to thank A. Tseytlin for comments and critically reading the draft of the paper. This work has been supported in part by an operating grant from the Natural Sciences and Engineering Research Council of Canada.

\appendix

\renewcommand{\theequation}{\Alph{section}.\arabic{equation}}

\sect{Appendix: Penrose limit of Pilch-Warner geometry}

In this Appendix we review the results of \cite{pw, myers} about Penrose limit of Pilch-Warner geometry at the IR fixed point and resulting superstring theory. 

Let us consider geodesics for which $\th=\a=0$ leading to the Lagrangian:
\eq{
\mathcal{L}= \lb(-\dot\t^2+\frac{4}{9}\tilde\psi^2\rb)
\label{a1.}
}
where\footnote{We note that $\tilde\psi$ differs from our $\psi$ defined in Section 3. In this appendix we use the notations of \cite{myers}.}
\eq{
\tilde\psi=\f+\b+\g .\notag
}
The light-cone coordinates which brings the metric into Brinkmann form after Penrose limit are defined as;
\eq{
u=\frac{1}{2E}\lb(\t+ \frac{2}{3}\tilde\psi\rb);\quad
v=-EL^2\Omega_0^2\lb(\t-\frac{2}{3}\tilde\psi\rb) \label{a.2}
}
where $E$ stands for the conserved energy associated with the Killing vector $\p_\t$. The conserved quantity associated with the other Killing vector $\p_\psi$ is the angular momentum $J$($=3/2\,E$). Penrose limit means that $L\to\infty$ should be taken and the fact that we consider geodesics for $\r=\th=\a=0$ means that they should be also scaled in appropriate way:
\eq{
\r=\frac{r}{L},\quad \th=\frac{y}{L}, \quad \a=\frac{w}{L} \label{a.3}.
}
Defining the angular coordinates $\hat{\f}$ and $\hat{\g}$ as
\eq{
\hat{\f}=\f-\frac{1}{3}\psi, \quad \hat{\g}=\g-\frac{2}{3}\psi ,
\label{a.4}
}
Penrose limit reduce the metric to the form:
\ml{
ds^2=2dx^+dx^--E^2\lb(r^2+w^2+4y^2\rb)d{x^+}^2+dr^2+r^2d\Omega_3^2+ dy^2  \\
+y^2d{\hat\f}^2+dw^2+w^2d{\hat\g}^2. \label{a.5}
}
The form fields then becomes
\al{
F_5=& -E(1+\star)dx^+\wedge\epsilon(\mathbb{E}^4) \notag \\
G_3=& -\sqrt{3}Ee^{-\b}dx^+\wedge(dy-iy^2d{\hat\f})\wedge (dw-iw^2d{\hat\g}). \label{a.6}
}
The bosonic part of the action with B-field turned on in directions $i=1,\cdots 4$ becomes:
\ml{
H_B=-\frac{1}{4\pi\a'}\int d^2\x \lb\{\sqrt{-g}g_{ab}\lb( 2\p_aX^+ \p_bX^-+A_{ij}X^iX^j\p_aX^+\p_bX^+ \right.\right. \\
 \left. +\p_aX^i\p_bX^i\rb) 
\left. -\sqrt{3}E\varepsilon^{ab}\lb(X^1\p_aX^+\p_bX^3 -X^2\p_aX^+\p_bX^4\rb)\rb\}
\label{a.7}
}
where
\eq{
A_{ij}=
\begin{pmatrix}
1 & 0 & 0 \\
0 & 1 & 0 \\
0 & 0 & 4 \\
\end{pmatrix} \label{a.8}
}
The standard light-cone gauge $X^-=\a' p^+\t+const~$ follows from the equations of motion for $X^-$ while the other equations of motion are
\al{
& \nabla^2X^1-4M^2X^1+\sqrt{3}M\p_\s X^3=0 \notag \\
& \nabla^2X^2-4M^2X^2-\sqrt{3}M\p_\s X^4=0 \notag \\
& \nabla^2X^3-M^2X^1-\sqrt{3}M\p_\s X^1=0 \label{a.9} \\
& \nabla^2X^4-M^2X^1+\sqrt{3}M\p_\s X^2=0 \notag \\
& \nabla^2X^p-M^2X^p=0, \notag 
}
where $M=E\a' p^+$ and the index $p$ stands for the directions unaffected by $B_2$.

Using the Virasoro constraints ($T_{ab}=0$) the action (\ref{a.7}) can be brought to the form
\ml{
S_B=-\frac{1}{4\pi\a'}\int d^2\x\lb\{-2\a' p^+\p_\t X^+-(\a' p^+)^2 A_{ij}X^iX^j+\e^{ab}\p_aX^i\p_bX^i \right. \\
\left. -2\sqrt{3}M(X^1\p_\s X^2-X^2\p_\s X^4)\rb\}.
\label{a.10}
}
Substituting the standard Fourier expansion for $X^i$
\eq{
X^i(\t\s)=\sum\limits_{i}C^i_n e^{i(\o_n+n\s)}
\notag
}
and assuming periodic boundary conditions we get the following solutions to the equations of motion
\ml{
X^p(\t\s)=\cos M\t x_0^p+\frac{\a'}{M}\sin M\t p_0^p \\
+ i\sqrt{\frac{\a'}{2}}\sum\limits_{i\neq 0}\frac{1}{\o_n}
\lb(\a_n^p e^{in\s}+\tilde\a_n^p e^{-in\s}\rb)e^{-i\o_n\t}
\label{a.11}
}
where:
\eq{
\o_{\pm n}=\pm\sqrt{n^2+M^2}, \quad (\a_n^p)^\dagger=\a_{-n}^p, \quad
(\tilde\a_n^p)^\dagger=\tilde\a_{-n}^p
\label{a.11a}
}
with
\eq{
[x_0^p,x_0^q]=\d^{pq}, \quad [\a_n^p,\a_m^q] =
[\tilde\a_n^p,\tilde\a_m^q] =\d_{n+m,0}^{pq}.
\notag 
}
The solutions in $p$-directions are
\ml{
X^A(\t,\s)=\cos 2M\t x_0^A+\frac{\a'}{2M}\sin 2M\t p_0^A \\
+i\sqrt{\frac{\a'}{2}}\sum\limits_{n\neq 0}\lb[\frac{1}{\o_n^+}
\lb(\b_n^A e^{in\s}+\tilde\b_n^A e^{-in\s}\rb)e^{-i\o_n^+\t} \right.\\
+\left.\frac{1}{\o_n^-}
\lb(\g_n^A e^{in\s}+\tilde\g_n^A e^{-in\s}\rb)e^{-i\o_n^-\t}
\rb] \label{a.12} 
}
\ml{
X^{A+2}(\t,\s)=\cos 2M\t x_0^{A+2}+\frac{\a'}{2M}\sin 2M\t p_0^{A+2} \\
+i\sqrt{\frac{\a'}{2}}\sum\limits_{n\neq 0}\lb[\frac{c_n^+}{\o_n^+}
\lb(\b_n^{A+2} e^{in\s}-\tilde\b_n^{A+2} e^{-in\s}\rb)e^{-i\o_n^+\t} \right.\\
+\left.\frac{c_n^-}{\o_n^-}
\lb(\g_n^{A+2} e^{in\s}-\tilde\g_n^{A+2} e^{-in\s}\rb)e^{-i\o_n^-\t}
\rb] \label{a.13} 
}
where $A=1,2$ and
\al{
& (\o_n^{\pm})^2=\frac{1}{2}\lb(2n^2+5M^2\pm \sqrt{12n^2M^2+9M^4}\rb)
\notag \\
& c_n^\pm=\frac{i}{2\sqrt{3}nM}\lb(-3M^2\pm\sqrt{12n^2M^2+9M^4}\rb);
\quad c_n^+c_n^-=1.
\label{a.13a}
}
The Hermitian conjugation and commutation relations are
\al{
& (\b_n^A)^\dagger=\b_{-n}^a, \quad (\g_n^A)^\dagger=\g_{-n}^a,
(\tilde\b_n^A)^\dagger=\tilde\b_{-n}^a, \quad (\tilde\g_n^A)^\dagger=\tilde\g_{-n}^a, \notag \\
& [x_0^A,x_0^B]_{PB}=[x_0^{A+2},x_0^{B+2}]_{PB}=\d^{AB}  
 \notag \\
& [\b_m^A,\b_n^B]_{PB}=[\tilde\b_m^A,\tilde\b_n^B]_{PB}=-i\o_m^+ \frac{c_m^-}{c_m^--c_m^+}\d_{n+m,0}\d^{AB}  
\notag \\
& [\g_m^a,\g_n^B]_{PB}=[\tilde\g_m^A,\tilde\g_n^B]_{PB}=-i\o_m^- \frac{c_m^+}{c_m^+-c_m^-}\d_{n+m,0}\d^{AB}   
\label{a.14}
}
The Virasoro constraints imposes $N=\tilde N$ with
\al{
& N=\sum\limits_{n\neq 0}n\lb[\frac{1}{\o_n}\a_{-n}^p\a_n^p + \frac{1}{\o_n^+}(1-{c_n^+}^2)\b_{-n}^A\b_n^A+
\frac{1}{\o_n^-}(1-{c_n^-}^2)\g_{-n}^A\g_n^A\rb] \notag \\
& \tilde N=\sum\limits_{n\neq 0}n\lb[\frac{1}{\o_n}\tilde\a_{-n}^p \tilde\a_n^p + \frac{1}{\o_n^+}(1-{c_n^+}^2)\tilde\b_{-n}^A \tilde\b_n^A+
\frac{1}{\o_n^-}(1-{c_n^-}^2)\tilde\g_{-n}^A\tilde\g_n^A\rb]
\label{a.15}
}
The Hamiltonian of the theory then becomes
\ml{
H_B=\frac{1}{2\a'}\lb[{\a'}^2p_0^ip_o^i+4M^2\sum\limits_{i=1,2}
x_0^ix_0^i+M^2\sum\limits_{i=3}^8x_0^ix_0^i\rb] \\
+\frac{1}{2}\sum\limits_{n\neq 0}\lb[\a_{-n}^p\a_n^p+
\tilde\a_{-n}^p\tilde\a_n^p+(1-{c_n^+}^2)(\b_{-n}^a\b_n^a+
\tilde\b_{-n}^a\tilde\b_n^a)\right. \\
\left.+(1-{c_n^-}^2)(\g_{-n}^a\g_n^a+\tilde\g_{-n}^a\tilde\g_n^a)\rb]
\label{a.16}
}
The quantization of the theory should be performed by replacing the Poisson brackets with commutators of the canonically normalized oscillators
\al{
& \a_n^p=\sqrt{\o_n}a_n^p, & & \a_{-n}^p=\sqrt{\o_n}\bar a_n^p,\notag \\
& \b_n^A=\sqrt{\o_n^+\frac{c_n^-}{c_n^--c_n^+}}b_n^A,
& &  \b_{-n}^A=\sqrt{\o_n^+\frac{c_n^-}{c_n^--c_n^+}}\bar b_n^A,\notag \\
& \g_n^A=\sqrt{\o_n^-\frac{c_n^+}{c_n^+-c_n^-}}c_n^A,
& &  \g_{-n}^A=\sqrt{\o_n^-\frac{c_n^+}{c_n^+-c_n^-}}\bar c_n^A, \notag \\
& a_0^i=\frac{1}{\sqrt{4M\a'}}(\a' p_0^i-2iMx_0^i),
& & \bar a_0^i=\frac{1}{\sqrt{4M\a'}}(\a' p_0^i+2iMx_0^i), (i=1,2) \notag \\
& a_0^i=\frac{1}{\sqrt{2M\a'}}(\a' p_0^i-iMx_0^i),
& & \bar a_0^i=\frac{1}{\sqrt{2M\a'}}(\a' p_0^i+iMx_0^i), (i=3,\dots 8). \notag
}
In this oscillator basis the Hamiltonian takes the form
\eq{
H=\Delta E+2M\sum\limits_{i=1,2}N_0^{(i)}+ M\sum\limits_{i=3}^8 
N_0^{(i)} +\sum\limits_{n>0}\lb(\o_nN_n^{(a)}+\o_n^+N_n^{(b)}+ \o_n^-N_n^{(c)} \rb) \label{a.18}
}
where $\Delta E$ is the zero point energy and the occupation numbers are defined in a standard way
\eq{
N_n^{(a)}=\bar a_n^pa_n^p+\tilde{\bar a}_n^p\tilde a_n^p, \quad
N_0^i=\bar a_0^ia_n^i \label{a19}
}
and similarly for $N_n^{(b)}$ and $N_n^{(c)}$. The frequencies $\o_n$ and $\o_n^\pm$ are given in (\ref{a.11a},~\ref{a.13}). The difference with the maximally supersymmetric case is encoded in the complicated form of the frequencies $\o_n^\pm$. 

We note that the same analysis can be applied to the case of point-like string (see section Conclusions). In this case $\r\to\r_0, \r'=0$ and the equations of motion becomes the same as (\ref{a.9}). Therefore, all the results and conclusions concerning quantum corrections to the classical energy are the same as in the above.

\end{document}